\documentclass[11pt]{JHEP3}
\usepackage{amsfonts}
\usepackage{amsmath}
\usepackage{graphicx}

\usepackage[all]{xy}

\newcommand{\dk}{\mathrm{d} k}
\newcommand{\CP}{\mathbb{CP}}
\newcommand{\dbar}{\bar\partial}
\newcommand{\cA}{\mathbf{A}}
\newcommand{\cB}{\mathbf{B}}
\newcommand{\ca}{{\mathbf a}}
\newcommand{\cb}{{\mathbf b}}
\newcommand{\cC}{{\mathbf{C}}}
\newcommand{\cc}{{\mathbf{c}}}
\newcommand{\cbC}{{\bar{\mathbf{C}}}}
\newcommand{\cbc}{{\bar{\mathbf{c}}}}
\newcommand{\PT}{{\mathbb{PT}}}
\newcommand{\tr}{\, \mathrm{tr}\;}

\newcommand{\dalpha}{\dot{\alpha}}
\newcommand{\dbeta}{\dot{\beta}}
\newcommand{\dgamma}{\dot{\gamma}}

\def\braket#1{\mathinner{\langle{#1}\rangle}}

\title{A quantization of twistor Yang-Mills theory through the background field method}
\author{Rutger Boels\\
The Mathematical Institute, University of Oxford \\ 24-29 St.
Giles, Oxford OX1 3LP, United Kingdom \\
${Rutger.Boels@maths.ox.ac.uk}$}


\abstract{Four dimensional Yang-Mills theory formulated through an action on twistor space has a larger gauge symmetry than the usual formulation, which in previous work was shown to allow a simple gauge transformation between text-book perturbation theory and the Cachazo-Svr\v{c}ek-Witten rules. In this paper we study non-supersymmetric twistor Yang-Mills theory at loop level using the background field method. For an appropriate partial quantum field gauge choice it is shown the calculation of the effective action is equivalent to (the twistor lift of) the calculation in ordinary Yang-Mills theory in the Chalmers and Siegel formulation to all orders in perturbation theory. A direct consequence is that the twistor version of Yang-Mills theory is just as renormalizable in this particular gauge. As applications an explicit calculation of the Yang-Mills beta function and some preliminary investigations into using the formalism to calculate S-matrix elements at loop level are presented. In principle the technique described in this paper generates consistent quantum completions of the CSW rules. However, by inherent limitations of the partial gauge choice employed here, this offers in its current form mainly simplifications for tree level forestry. The method is expected to be applicable to a wide class of four dimensional gauge theories.}

\begin{document}

\section{Introduction}
Even more than 50 years after its introduction, Yang-Mills theories in four space-time dimensions continue to be a fascinating and rich area of research. Applications range from the very physical in the standard model to the very theoretical in the study of geometric invariants of (four) manifolds. However, it is fair to say that they are, in some sense, not very well understood. Their non-perturbative behaviour for instance is not under analytic control. The only known exceptions to this involve either less space-time dimensions or supersymmetry. In other words, in situations where there is an extra underlying symmetry to exploit. In addition, also the perturbative behaviour of Yang-Mills theory for the calculation of scattering amplitudes is a very active area of research still because of its computational complexity when using standard methods even at tree level, especially when large numbers of external particles are involved. That is not to say these scattering amplitudes are not interesting: in order to discover new physics at LHC for instance one needs a good quantitative control of the `old' physics contained in these amplitudes. However, for the strong force for instance the coupling constant is not parametrically small in the region of interest and one needs to calculate loop corrections which is prohibitively complex when using ordinary Feynman diagrams. Then one has to resort to other calculational methods. 

However, there are several results in the literature which show that even ordinary perturbation theory based on the non-supersymmetric space-time Yang-Mills action misses part of an underlying structure of the theory. The first of these is the classic result of Parke and Taylor \cite{parktaylor} that a particular class of amplitudes have a simple expression at tree level. These are the amplitudes which involve two gluons of one, and arbitrarily many gluons of the opposite helicity. This result was related to a construction on twistor space by Nair \cite{nair}. The connection between perturbative amplitudes and twistor space was further elaborated upon by Witten \cite{witten}, who showed that it can be extended at least to all tree level amplitudes. In addition, he also speculated that these results could be derived from an underlying topological string theory which should be equivalent to $\mathcal{N}=4$ super Yang-Mills theory. Note that Witten's twistor string proposal can be understood as an attempt to answer the question \emph{why} some results in perturbative Yang-Mills theory like the Parke-Taylor amplitude are so simple.

This inspired a lot of activity in the last few years, which up to now has mainly focused on obtaining new calculational techniques for scattering amplitudes. Two outstanding developments here are the recursive techniques of Britto, Cachazo, Feng and Witten \cite{BCFW} and the Feynman-like rules of Cachazo, Svr\v{c}ek and Witten \cite{CSW}. Both these techniques rearrange the ordinary perturbative expansion of Yang-Mills theories into something much more simple. A natural physical explanation for this possibility is that there is some underlying symmetry in the problem which is not manifest in the usual Yang-Mills Lagrangian. In previous work \cite{lionel1, us, usII} this natural conjecture was confirmed by constructing explicitly an action on twistor space with a (linear) gauge symmetry which is larger than the ordinary gauge symmetry. In \cite{us, usII} it was shown that both the CSW rules and text-book perturbation theory can be obtained as Feynman rules for the twistor action\footnote{We expect that the BCFW rules can be derived from an action written on ambi-twistor space \cite{lioneldavid} by direct derivation  of the intriguing twistor diagram formalism of \cite{hodges} from the ambitwistor action. An alternative avenue of attack is the technique of \cite{leenair}. Both these approaches will involve a gauge choice similar to the CSW gauge.}. 

In related work, Mansfield \cite{mansfield} obtained an action which reproduces the CSW rules on-shell by a non-linear and non-local canonical field transformation from the light-cone formulation of four-dimensional Yang-Mills. It is expected that that field transformation is exactly the space-time equivalent of the twistor gauge transformation of \cite{usII}. In this article we provide some evidence for this claim; a full treatment will appear elsewhere. In more physical terms, from a space-time point of view we conjecture the twistor action provides the right auxiliary fields to linearise a specific non-linear space-time symmetry and is in that sense a `superfield' formulation\footnote{It has recently been pointed out to the author by K. Stelle that this can be made more precise. The reader averse to the words `twistor space' is therefore encouraged to read `non-supersymmetric Lorentzian harmonic superspace' \cite{sokatchev} instead at occurring places.}.

A remaining open problem in all this is extending the analysis to loop level. Ordinary text-book perturbation theory is fine, but in straightforwardly applying the CSW rules to loop level \cite{qmul}, one encounters problems as at one loop the diagrams generically only reproduce the cut-constructible pieces of amplitudes. This is a problem in particular for non-supersymmetric gauge theories, where amplitudes are known to involve more than just cut-constructible parts. Recently some interesting light was shed on this in \cite{BSTrecent}, although a full understanding is still lacking. From the view of twistor Yang-Mills theory however, we seem to have a gauge theory which interpolates between a well-defined and a not-so-well-defined perturbation theory. In this article the twistor action will be quantised in a gauge close to, but not equivalent to space-time gauge where all the divergences are simply four-dimensional. In this gauge the regularization and renormalization properties will be shown to reduce to standard space-time problems, which can be resolved by standard techniques. This is part of the main message of this paper: standard four dimensional field theory techniques extend to Yang-Mills theories on twistor space. 

This article is structured as follows: We will begin by giving a brief review and clarification of the twistor action approach to Yang-Mills theory and point out some of its more salient features. After this the background field method will be set up, with the quantum field in the background field version of the space-time gauge. This will lead to specific Feynman rules which can then be identified with the space-time Yang-Mills rules derived from the Chalmers and Siegel action. Put differently, the calculation shows that the space-time quantum effective action can be lifted to twistor space by a simple lifting prescription. In particular, the $\beta$ function calculation, performed explicitly in an appendix for the Chalmers and Siegel action, lifts directly, as well as renormalizability arguments. The next section contains some investigations into using the formalism to calculate scattering amplitudes. In an appendix an alternative gauge for obtaining CSW rules is constructed. 

In this article dotted and undotted Greek letters from the beginning of the alphabet indicate spinor indices. Our spinor conventions are $\omega \cdot \lambda = \omega^\alpha \lambda_{\alpha} = \omega^\alpha \lambda^\beta \epsilon_{\beta \alpha}$, $\pi \cdot \mu = \pi_{\dalpha} \mu^{\dalpha} = \pi_{\dalpha} \mu_{\dbeta} \epsilon^{\dbeta \dalpha}$. We normalise the isomorphism between the cotangent bundle and the spin-bundles such that $g_{\mu \nu} = \frac{1}{2} \epsilon_{\alpha \beta} \epsilon_{\dalpha \dbeta}$, where (symbolically) $\mu = \alpha\dalpha$, $\nu = \beta \dbeta$. Furthermore, fields on twistor space are denoted by Roman symbols, while space-time fields are bold. Quantum and background fields will be denoted by lower and upper case letters respectively. Finally, we normalise the natural volume-form on a $\CP^1$ such that it includes a factor of $\frac{1}{2 \pi i}$.

\section{Twistor Yang-Mills theory}
In this section the twistor formulation of Yang-Mills theory will be reviewed and clarified. Although this is not immediately obvious from the exposition here, what is discussed in this section is a Euclidean, \emph{off-shell} version of the Penrose-Ward correspondence and the interested reader is referred to \cite{woodhouse} for an introduction to twistor space useful from the point of view of this paper. 

\subsection{Some twistor geometry}
As usual, four-dimensional space-time arises in the twistor program as the space of holomorphic lines embedded in $\CP^3$'. The prime indicates the removal of a $\CP^1$, which is necessary mathematically to obtain interesting cohomology as $\CP^3$ is compact, and physically to obtain a notion of a point at `infinity' which is needed to define scattering amplitudes. Consider homogeneous coordinates $(\omega^{\alpha}, \pi_{\dalpha})$ for a point in $\CP^3$' where $\alpha$ and $\dalpha$ run from $1$ to $2$. Then a holomorphic line corresponds to an embedding equation
\begin{equation}
\omega^{\alpha} = x^{\alpha \dalpha} \pi_{\dalpha}.
\end{equation}
Note that this equation makes sense since the symmetry group of twistor space contains naturally two $SL(2,\mathbb{C})$ subgroups which can and will be identified with the chiral components of the complexified Lorentz group. In order for this equation to be solved for $x$ a reality condition is needed. In this paper we will be interested in Euclidean signature for which Euclidean spinor conjugation is needed, 
\begin{equation}
\widehat{\left(\begin{array}{c} \pi_1 \\ \pi_2 \end{array} \right)} = \left(\begin{array}{c} - \bar{\pi}_2 \\ \bar{\pi}_1 \end{array} \right).
\end{equation}
The main difference to the Lorentzian conjugation is the absence of the application of the parity operator which interchanges spin bundles. In Euclidean signature there is a \emph{unique} point $x$ associated to every pair $\omega, \pi$,
\begin{equation}
x^{\alpha \dalpha} = \left(\frac{\omega^{\alpha} \hat{\pi}^{\dalpha} - \hat{\omega}^{\alpha} \pi^{\dalpha} }{\pi \hat{\pi}} \right)  
\end{equation}
this equation exposes twistor space for Euclidean signature space-time as a $CP^1$ fibre-bundle over space with the above equation furnishing the needed projection, which will be denoted by $p$. 

In the following explicit coordinates $x^{\alpha \dalpha}, \pi_{\dalpha}, \hat{\pi}_{\dalpha}$ will be used to parametrize the twistor space. This choice leads to a basis of anti-holomorphic one-forms,
\begin{equation}
\bar{e}_0 = \frac{\hat{\pi}^{\dalpha} d \hat{\pi}_{\dalpha}}{(\pi \hat{\pi})^2} \quad  \bar{e}^\alpha = \frac{ dx^{\alpha \dalpha} \hat{\pi}_{\dalpha}}{(\pi \hat{\pi})}
\end{equation}
which is naturally dual to a set of $(0,1)$ vectors,
\begin{equation}
\dbar_{0} \equiv (\pi \hat{\pi}) \pi_{\dalpha} \frac{\partial}{\partial \hat{\pi}_{\dalpha}} \quad \dbar_{\alpha} \equiv  \pi^{\dalpha} \frac{\partial}{\partial x^{\alpha \dalpha}}.
\end{equation}
With the above basis of one-forms, one-form fields $A$ can be expanded as
\begin{equation}
A =  \bar{e}^\alpha A_\alpha+  \bar{e}^0 A_0
\end{equation}
Note that $A_{\alpha}$ and $A_0$ have holomorphic weight $+1$ and $+2$ respectively compared to the original weight of $A$. In the following the word `weight' will always refer to holomorphic weight. 

\subsection{Lifting fields to twistor space}
One of the successes of the twistor program has always been the fact that on-shell fields, including self-dual fields, correspond to certain cohomology classes on twistor space. However, here we will be interested in lifting off-shell fields from space-time to twistor space, clarifying the procedure first employed in \cite{lionel1}. The objective is to lift Yang-Mills theory, such as captured in the Chalmers and Siegel action
\begin{equation}
S_{\textrm{CS}} = \tr \int d^4\!x\, \frac{1}{2}  \cB_{\dalpha \dbeta} F^{\dalpha \dbeta}[\cA] - \frac{1}{4} \tr \int d^4x\, \cB_{\dalpha \dbeta} \cB^{\dalpha \dbeta}
\end{equation}
from Euclidean space-time to twistor space. First of all, the self-dual two form $\cB$ lifts as
\begin{equation}
\cB_{\dalpha \dbeta} = \int_{\CP^1}\dk\, H^{-1} B_0 H \pi_{\dalpha} \pi_{\dbeta} 
\end{equation}
Here $\dk$ is the natural weightless volume-form on $\CP^1$ and $B_0$ is the zeroth component of a weight minus 4, anti-holomorphic form. Anti-holomorphic $n$-forms will be denoted by $(0,n)$. $H$ is a holomorphic frame of the gauge bundle over $p^{-1}(x)$ such that the covariant derivative of it vanishes on the sphere, $\dbar_A H|_{p^{-1}(x)}=0$. This covariant derivative involves a connection of the Riemann sphere which is always trivial in perturbation theory\footnote{There are of course non-trivial connections on the $\CP^1$ with vanishing curvature. As we will only be interested in perturbation theory in this paper these are avoided by the smallness assumption on $A$.}. Denote this connection (or more precisely, it's $(0,1)$ part) as $A_0$. With this input the Chalmers and Siegel action becomes,
\begin{align}\label{eq:twacthalfway}
S_{CS} = \tr \int d^4\!x \dk\, & \frac{1}{2}  B_0  H F_{\dalpha \dbeta}(x)[\cA] H^{-1} \pi^{\dalpha} \pi^{\dbeta} - \nonumber \\
& \frac{1}{4} \tr \int d^4x \dk_1 \dk_2\, H_1^{-1} B^{(1)}_0 H_1 H_2^{-1} B^{(2)}_0 H_2 (\pi_{1} \pi_2)^2
\end{align}
Now we would like to lift the gauge field $\cA_{\mu}$ to twistor space. We already know from basic twistor theory (see \cite{woodhouse} for instance) that it is represented by a weight $0$ $(0,1)$ form, say $A$, which is the pull-back of the space-time connection to the twistor space. If this form is $\dbar$ closed, then it corresponds to an anti-self-dual connection on space-time. The curvature of this form naturally splits into a curvature tensor involving only space-directions, say $F_{\alpha \beta} = [\dbar_\alpha + A_\alpha, \dbar_\beta + A_\beta]$ and two curvature tensors with one leg along the fibre $F_{0 \alpha} = [\dbar_0 + A_0, \dbar_\alpha + A_\alpha] $. Since there is no physical interpretation of the latter curvature and since we want $A$ to be just the pull-back of the physical space-time connection, it will be set to zero: we will study gauge connections on twistor space such that $F_{0 \alpha}$ vanishes. As this is a weight $3$ operator, the condition it vanishes can be added to the action with a Lagrange multiplier of weight $-3$,
\begin{equation}\label{eq:accons}
S \,+\!= \frac{1}{2} \int d^4\!x \dk\, B^{\alpha} F_{0 \alpha}
\end{equation}
This constraint also gives a neat way of lifting the vector potential $A$: For calculational ease, let $F_{0\alpha}$ act on a field in the fundamental,
\begin{equation}
F_{0\alpha} \phi = [(\dbar_A)_{0}, (\dbar_A)_{\alpha}] \phi
\end{equation}
Since $A_0$ is pure gauge, it can be always be gauged away. This is implemented through the use of the holomorphic frames $H$. We arrive at
\begin{equation}
F_{0\alpha} \phi = \dbar_{0}\left(H^{-1} (\dbar_\alpha + A_{\alpha}) H \right) H^{-1} \phi 
\end{equation}
The constraint sets this quantity to be zero. The quantity in brackets is then a holomorphic function of weight one and therefore
\begin{equation}\label{eq:vectcohom}
H^{-1}\left( \dbar_\alpha + A_{\alpha}  \right)H = \cA_{\alpha \dalpha}(x) \pi^{\dalpha} 
\end{equation} 
for some vector field $\cA_{\alpha \dalpha}$ which is only a function of $x$. This can easily be inverted to give 
\begin{equation}\label{eq:eqforpentranA}
\cA_{\alpha \dalpha}(x) =  \int \dk H^{-1} \left(\dbar_\alpha + A_{\alpha}  \right)H  \frac{\hat{\pi}_{\dalpha} }{\pi \hat{\pi}}
\end{equation}
In the same way as above, we easily derive
\begin{equation}
\epsilon^{\alpha \beta} F_{\alpha \beta}(x, \pi) = H F[\cA]_{\dalpha \dbeta}(x) H^{-1} \pi^{\dalpha} \pi^{\dbeta} 
\end{equation}
With this expression taken into account and the constraint in (\ref{eq:accons}) added, the action (\ref{eq:twacthalfway}) becomes
\begin{align}\label{eq:twymactionXXX}
S =  \frac{1}{2} & \int d^4\!x \dk B_0 \left(\dbar^{\alpha} A_{\alpha} + g A^{\alpha} A_{\alpha} \right) + B^{\alpha} \left(\dbar_{\beta} A_0 - \dbar_0 A_\beta + g[A_\beta,A_0]\right) \nonumber\\
& - \frac{1}{4} \int d^4\!x \dk_1 \dk_2 H^{-1}_1 B^0(\pi_1) H_1 H^{-1}_2 B^0(\pi_2) H_2 (\pi_1 \pi_2)^2 (\pi_{1} \pi_2)^2
\end{align}

\noindent This action has a clear geometrical expression on twistor space, as the fields $A_0$, $A_\alpha$, $B_0$ and $B_\alpha$ can naturally be combined into anti-holomorphic one forms $A$ and $B$ of weight $0$ and $-4$ respectively,
\begin{align}\label{eq:twymaction}
S[A,B] = \frac{1}{2} &\int_{\PT} D^3Z \wedge B \wedge \left(\dbar A + A \wedge A \right)   -\nonumber \\ & \frac{1}{4}  \int_{\PT\times_\mathbb{M} \PT} 
\tr H_1^{-1}B_1H_1\wedge H_2^{-1}B_2H_2\wedge  D^3Z_1\wedge D^3Z_2
\end{align}
where $\PT$ is projective twistor space $\CP^3$' with a space-time point removed, $\PT\times_\mathbb{M} \PT=\{(Z_1,Z_2)\in\PT\times\PT|p(Z_1)=p(Z_2)\}$ with $p$ the projection map and subscript 1 or 2 denotes dependence on $Z_1$ or $Z_2$. This form of the action is the restriction to the $\mathcal{N}=0$ fields of the $\mathcal{N} =4$ form of the twistor Yang-Mills action as studied in \cite{us, usII} and this action appeared in this form first in \cite{lionel1}, although there only the twistor lift of $B$ was studied.

\subsection{Gauge invariances}
The twistor action is invariant under
\begin{equation}\label{eq:gaugeinv_reg}
A \rightarrow A+ \dbar_A \chi \quad B \rightarrow B+ g [B,\chi] + \dbar_A \omega
\end{equation}
which can either be verified explicitly, or inferred directly from the $\mathcal{N}=4$ formulation. Here $\chi$ and $\omega$ are functions on projective twistor space of weight $0$ and $-4$ respectively. This apparently innocuous remark is actually very important: $\chi$ is a function of $6$ real variables, instead of the usual $4$. Hence this formulation of Yang-Mills theory has \emph{more} gauge symmetry than the usual one. It is the existence of this symmetry which is the underlying physical reason for the existence of MHV methods in general. For instance, the Parke-Taylor formula (and its supersymmetric analogs) can be viewed as a consequence of this symmetry, as shown in \cite{usII}. 

Note that in the `lifting picture' the gauge symmetries have quite dissimilar origins: the gauge symmetry in $A$ is in effect space-time gauge symmetry + gauge symmetry on the $P_1$ for $A_0$ and $A_\alpha$ separately, but since $F_{0\alpha}$ vanishes, this can be enlarged to the symmetry group above. The gauge invariance in $B_0$ is a consequence of some leeway in the `lifting' formula. The full gauge symmetry in $B$ is actually the most interesting since it is a direct consequence of `quantising with constraints': when one quantises a theory with constraints, in a real sense also the momentum conjugate to the constraint must be eliminated. This is familiar from the usual setup of gauge theories as explained in, for instance, \cite{itzyksonzuber} as this leads to the usual gauge fixing procedure. The gauge symmetry associated to $B$ in the above action is generated by $F_{0 \alpha}$ and gauge fixing this in the usual manner is therefore the right procedure by the same calculation as for the ordinary gauge symmetry. 

The same procedure as employed here for the gluon can be extended readily to spin-0 and spin-$\frac{1}{2}$ fields in an obvious fashion. The lifting of these fields will entail separate new gauge invariances. One wants to require full gauge invariance with respect to these as this is necessary to allow an invertible field transformation. Requiring this leads to additional towers of $B^2$ like terms in the action through the Noether procedure. These will generate, in gauge which will  be discussed in the next section, additional towers of MHV-like vertices and amplitudes as was shown in \cite{usII}. 

\subsection{Quantization generalities}
Standard path integral quantization of the twistor version of Yang-Mills theory will involve a gauge choice as this is needed to invert the kinetic operator. In this section some some general aspects of this will be explored. Before the consequences of gauge invariance will be explored it is perhaps useful to point out some features which are apparent in this form of the action. First of all the mass-dimension of the fields as counted by space-time derivatives are rather odd, as they are
\begin{equation}
[A_0]  = 0 \quad [A_\alpha] = 1 \quad [B_0] = 2 \quad [B_A] = 2 
\end{equation}
The vanishing dimension of the field $A_0$ is worrying as one expects this will lead to infinite series of possible counterterms. Very naively, this action is not power-counting renormalizable! However, it is also clear the quadratic part of this action is definitely not canonical and involves the fibre coordinates, which makes power counting non-standard. This is related to a second comment: although an action on a six-dimensional space is studied for a four-dimensional theory, one does not expect Kaluza-Klein modes, since that argument requires a \emph{canonical} six-dimensional kinetic term. Thirdly, this action is non-local. However, the non-locality is restricted to the $\CP^1$ fibre, not on space-time, so this is not a problem provided this remains true in perturbation theory. 

\subsubsection{Space-time gauge}
The gauge invariance of the action can be exploited in multiple ways. The same techniques as in \cite{us, usII} also apply here. In particular, when restricted to a gauge in which
\begin{equation}\label{eq:spacetimegauge_reg}
\dbar_0^\dagger a_0 =0 \quad \quad \dbar_0^\dagger b_0 =0
\end{equation}
the above action reduces down to the Chalmers and Siegel form of the non-supersymmetric Yang-Mills action, which is perturbatively equivalent to the usual one. This gauge will be referred to as `space-time gauge'. Note that it does not fix the complete gauge symmetry contained in (\ref{eq:gaugeinv_reg}). The residual gauge symmetry is exactly the usual space-time one, which of course has to be fixed further in order to quantise the theory. Any ordinary gauge choice will do for this. The calculation in \cite{us} will not be reproduced here, as it is an obvious specialisation of the argument in section \ref{sec:bgtwym}. In effect this article extends the above observation to a general class of gauges. 

As the action reduces to the usual space-time action in a particular partial gauge (and the associated ghosts can be shown to decouple), it should be obvious that path integral quantization of this theory is perfectly fine in this gauge: here it is just the usual quantization and \emph{every} perturbative calculation can therefore be reproduced using the twistor action to all orders in perturbation theory. The real question is if the same holds true in other gauges. Put differently, the question is whether the extra symmetry found in our formulation of Yang-Mills at tree level is in some way anomalous. More precisely, the question is whether physical quantities like scattering amplitudes are invariant under the extra gauge symmetry. 

\subsubsection{On quantization in CSW gauge}
In addition, as the Queen Mary group has shown \cite{qmul}, the most straightforward application of the Cachazo-Svrcek-Witten \cite{CSW} rules already does calculate some loop level effects: these rules can be used at the one-loop level to calculate the cut-constructible parts of scattering amplitudes. As shown explicitly in \cite{usII} these CSW rules can be derived directly from the twistor action by changing gauge to the axial (space-cone) like gauge
\begin{equation}\label{eq:CSWgauge}
\eta^{\alpha} A_{\alpha} =0 \quad  \eta^{\alpha} B_{\alpha} =0
\end{equation}
which we will call CSW gauge. Here $\eta$ is an arbitrary spinor, normalised such that $\eta \hat{\eta} =1$. That the CSW rules can be derived in this way is a possibility which is clear from the original article\cite{CSW}: the twistor action in the $\mathcal{N}=4$ case can also be obtained as the reduction to single trace-terms of the conjectured effective action of Witten's twistor-string theory. Hence the original derivation of the rules applies here and this is fleshed out in \cite{usII}. Therefore there are already two gauges in which our twistor formulation makes some sense at loop level: we have full consistency for the space-time gauge and partial consistency (at least self-consistency) for the CSW gauge.

However, the CSW rules do not calculate full amplitudes in non-supersymmetric Yang-Mills theories. A particular example of this drawback are the amplitudes with all helicities equal. These are zero at tree level for all pure Yang-Mills theories and vanish at loop level for supersymmetric ones. If one picks a preferred helicity for the MHV amplitudes used in the CSW rules, say the set which is `mostly minus', then it is obvious there is no one-loop diagram which has only minus on the external legs, although there are diagrams for `only' plus.\footnote{In the recent paper \cite{BSTrecent} these diagrams were shown to give the correct amplitude in the four point case. Since these diagrams are actually equivalent to Yang-Mills in light-cone gauge, this is expected.} In the approach of Mansfield these missing diagrams are thought to be generated by a Jacobian factor which appears if the canonical transformation is regulated properly. This scenario was supported by the calculations in \cite{BSTrecent} where a modified, non-canonical transformation was employed. This however invalidates the equivalence theorem and calculating amplitudes should then be done by combining contributions from many different sources.

In the twistor action approach a non-trivial Jacobian might translate to a partly anomalous gauge symmetry: the lifting formulae only guarantee space-time gauge symmetry, but nothing beyond that. A more mundane explanation would be the fact that there are regularization issues when one tries to quantise the twistor action at loop level in the CSW gauge. The most natural regularization treats the six dimensional twistor space as $(4-2\epsilon) + 2$: one performs a half-Fourier transform w.r.t. the space-time directions and employs dimensional regularization there. However, this already encounters several problems. It is well-known in four dimensions for instance that axial gauges need careful regulating \cite{leibbrandt} since Fourier transforms of 
\begin{equation}
\sim \frac{1}{\eta^{\alpha} \pi^{\dalpha} p_{\alpha \dalpha} }
\end{equation}
are ill-defined and a pole prescription is necessary. This problem afflicts the twistor action, since in CSW gauge the propagators can be shown to behave like $:A_0 B_0: = \frac{\delta(\eta \pi p)}{p^2}$, and this delta function is obtained as
\begin{equation}
\delta(\eta \pi p) \sim \dbar_0 \frac{1}{\eta \pi p }
\end{equation} 
A drawback of implementing the Mandelstam-Leibbrandt regulating prescription is that it leads to very large algebraic complexity. In addition, there is the usual problem using chiral indices in any dimensional regularization scheme. In a supersymmetric theory one would rely on dimensional reduction to keep all the spinor algebra in four dimensions. This is however known to be in grave danger of being inconsistent (non-unitary) in non-supersymmetric theories which leads to the inclusion of $\epsilon$-scalars. Furthermore, as we are regulating a gauge theory, Pauli-Villars is flawed as this would break gauge invariance, apart from the non-generic form mass-terms on twistor space take. 

In appendix \ref{app:difgaugcond} a gauge condition equivalent to the CSW gauge at tree level is constructed using 't Hooft's trick which has slightly better loop behaviour, but which remains problematic for basically the same reasons indicated above.

\section{Background field method for twistor Yang-Mills}
\label{sec:bgtwym}
The existence of the CSW rules and their loop-level application suggests a trick: If one is given a gauge theory with two gauges, one of which is well-defined at loop level and one which makes results most transparent, the obvious game to play is the quantization of this theory using the background field method. Within this method the quantum effective action is calculated by integrating over the quantum field, and when interpreted as the generating functional of 1PI diagrams, this can be used to calculate S-matrix elements for the background fields. As pointed out in \cite{abbottgrisaru} in the context of four-dimensional field theory, one can put the background field into a different gauge than the quantum field. As our action is very close to the four-dimensional one, it is reasonable to expect the same trick can be performed here. Hence in this paper the twistor Yang-Mills action will be studied with the quantum field in the background field version of the space-time gauge. Then some properties of the S-matrix will be studied if the background field is put into the CSW gauge. A very important product of the analysis is the study of renormalization properties of our action. Actually, this was the original reason to study background field formulations, both in the literature as for us. In particular a background field calculation will also easily yield the $\beta$ function of the theory, and it is this purpose for which the background field method is usually employed.

Since the twistor action reduces to the Chalmers and Siegel formulation of Yang-Mills theory in ordinary space-time gauge, as a suitable warm-up for the calculation about to be presented one should therefore first treat that action in the background field formalism. We refer the reader to appendix \ref{app:modelcalc} for the details of that calculation. Below the same procedure is performed for twistor Yang-Mills, which is shown to reduce the calculation down to the space-time one worked out in the appendix. 

Begin by splitting the fields $\tilde{A}$ and $\tilde{B}$ into a background and quantum part
\begin{equation}\label{eq:splitfieldsTWYM}
\tilde{A} = A + a \quad
\tilde{B} = B + b
\end{equation}
indicated by capital and lower case letters respectively. The action will be expanded in quantum fields, ignoring the terms linear in the quantum field as they will not contribute to the quantum effective action which is the generating functional of 1PI diagrams. Subsequently integrating out the quantum field requires a gauge choice. As the action is invariant under
\begin{eqnarray}
\tilde{A} &\rightarrow& \tilde{A} + \dbar_{\tilde{A}} \chi \\
\tilde{B} &\rightarrow& B + [\tilde{B}, \chi] + \dbar_A \omega
\end{eqnarray}
there are two obvious choices one can make for the symmetry transformations of quantum and background field:
\begin{align}\label{eq:gaugesym1}
A &\rightarrow A + \dbar_{A} \chi & A &\rightarrow A\nonumber  \\
B &\rightarrow B +  [B, \chi] + \dbar_{ A} \omega & B &\rightarrow B \nonumber \\
a &\rightarrow a +  [a, \chi] & a &\rightarrow a + \dbar_{A+a} \nonumber \chi\\
b &\rightarrow b +  [b, \chi] + [a,\omega]& b &\rightarrow b + [B+b, \chi]+ \dbar_{A+a} \omega
\end{align}
The objective is to completely fix the second symmetry, while keeping the first (referred to as background gauge symmetry). As the quantum fields transform in the adjoint under the background symmetry, writing elliptic gauge fixing conditions such as Lorenz gauge and the background field version of space-time gauge (\ref{eq:spacetimegauge_reg}) requires one to lift the derivatives to covariant derivatives. In addition, although its almost immaterial to our calculation, one should promote the Lagrange multiplier to transform in the adjoint of the background symmetry. Hence the ghost and anti-ghost will also transform in that adjoint by the usual BRST symmetry. 

\subsection{Gauge fixing the background field}
The background version of space-time gauge (\ref{eq:spacetimegauge_reg}) reads:
\begin{equation}\label{eq:spacetimegauge_BG}
\dbar_0^\dagger(H[A_0] a_0 H[A_0]^{-1}) =0 \quad \quad \dbar_0^\dagger(H[A_0] b_0 H[A_0]^{-1}) =0
\end{equation}
Note that this gauge condition involves a choice of metric on a $\CP^1$. These conditions are solvable since Yang-Mills connections on a $\CP^1$ are trivial. We can therefore transform to the frame where $A_0$ is zero, solve the equations and the transform back. Since both $a_0$ and $b_0$ are part of a one form on $\CP^1$ they are automatically $\dbar$ closed and by the gauge condition co-closed in the frame where $A_0$ is zero. They are therefore harmonic. As $a_0$ has weight zero and $b_0$ has weight $-4$, there are no non-trivial harmonic forms $a_0$ and there is a two dimensional space of harmonic forms $b_0$ by a standard cohomology calculation. Hence we obtain
\begin{equation}
a_0 = 0 \quad b_0 = \frac{3 H \cb_{\dalpha \dbeta}(x) H^{-1} \hat{\pi}^{\dalpha} \hat{\pi}^{\dbeta}  }{(\pi \hat{\pi})^2} 
\end{equation}
where $H$ are the holomorphic frames encountered before and we apologise for the appearance of a normalisation factor proportional to $(2 \pi i)$. Note that these frames are functionals of the background field $A_0(x,\pi)$. This solution can be put back into the action. The field $b_\alpha$ is now a Lagrange multiplier for a very simple condition,
\begin{equation}
(\dbar_0 +A_0) a_\alpha = 0.
\end{equation}
This can be solved in the same way as before, which yields
\begin{equation}
a_\alpha = H \ca_{\alpha \dalpha}(x) H^{-1} \pi^{\dalpha}.
\end{equation}
In the original derivation \cite{us} the ghosts which came from the space-time gauge fixing decoupled. In the present context, due to the presence of a coupling to the background field in both the gauge fixing condition \ref{eq:spacetimegauge_BG} and the symmetry transformations (of the symmetry we are trying to fix!), the argument is slightly more convoluted. First note that we can still argue that the coupling of the \emph{quantum} fields to ghosts is off-diagonal, so that these quantum fields can safely be ignored. That leaves a possible one-loop contribution to the effective action, since the diagonal part of the ghost action takes the form
\begin{equation}\label{eq:ghostdecop}
\sim \bar{c}(x,\pi) (\dbar_0 +A_0)^{\dagger}(\dbar_0 +A_0)c(x,\pi)
\end{equation}
The ghosts in this action however do not have a space-time kinetic term. This leads to a contribution to the effective action which seems to diverge wildly since it is proportional to 
\begin{equation}
\int d^4p = \delta^{(4)}(0).
\end{equation}
However, it is well known\footnote{but initially not to the author. A discussion can be found for instance in \cite{zinnjustin}.} that contributions like this vanish in dimensional regularization, just like tad-poles. Hence the ghosts which came from fixing space-time gauge can safely be ignored in perturbation theory, as long as dimensional regularization is employed. See also the discussion in section \ref{sec:towardssmatrix} for a second reason why this factor can safely be ignored.

\subsection{Reduction to space-time fields}
At this point it is clear that the quantum fields only live on space-time and the calculation starts to become equivalent to the calculation in \ref{app:modelcalc}. In particular the fields in a loop are now a standard four dimensional vector field and self-dual tensor field. This argument therefore neatly avoids any regularization problems special to the twistor space formulation. The quantum self-dual tensor field $\cb_{\dalpha \dbeta}$ can be integrated out to yield
\begin{align}\label{eq:YMbackgroundaction}
S[A,B, a,b] & = S[A,B] + \frac{1}{4} \int d^4x F_{\dalpha \dbeta}[\ca] F^{\dalpha \dbeta}[\ca]+ \nonumber \\
& \frac{3 g}{4 } \int d^4x F_{\dalpha \dbeta}[\ca] \int dk\left[ \right(H^{-1} (\dbar^\alpha + A^{\alpha}) H \left), \ca_{\alpha \dgamma}  \right] \frac{\pi^{\dgamma} \hat{\pi}^{\dalpha} \hat{\pi}^{\dbeta}}{(\pi \hat{\pi})^2} + \nonumber \\
&\frac{9 g^2}{16} \int d^4x \left[\int dk\left[ \left(H^{-1} (\dbar^\alpha + A^{\alpha}) H \right), \ca_{\alpha \dgamma}  \right] \frac{\pi^{\dgamma} \hat{\pi}^{\dalpha} \hat{\pi}^{\dbeta}}{(\pi \hat{\pi})^2} \right]^2 \nonumber  + \\
& \frac{3 g}{4} \int d^4x \ca^\alpha_{\dalpha} \ca_{\alpha \dbeta} \int dk H^{-1} B_0 H \pi^{\dalpha} \pi^{\dalpha}
\end{align}
As a final step one can now use (\ref{eq:vectcohom}) to argue that the coupling of quantum to background fields is the same as in the Chalmers and Siegel Lagrangian (\ref{eq:backgroundChalSieg}), since with this argument in hand the $\pi$ integrals can simply be performed. Actually, this is the calculation which lead to the derivation of (\ref{eq:vectcohom}) in the first place. 

The kinetic term for the quantum fields is just the ordinary Yang-Mills one up to a non-perturbative term. Therefore, dimensional regularization can be employed as usual. We will employ the original 't Hooft-Veltman \cite{thooftveltman} scheme which keeps the fields outside the loop in 4 dimensions. A second remark is on the structure of the vertices: every background field vertex involves an infinite amount of $A_0$ fields through the holomorphic frames. Note that this is permitted since the mass dimension of $A_0$ is zero. As a final remark note that this theory is expected to diverge no worse than Yang-Mills theory as the quantum field is simply a gluon: in this particular background gauge twistor Yang-Mills theory is therefore power counting renormalizable. This in contrast to naive power counting based on the mass dimension of the $a_0$ field.

Another way of summarising the above observation is that in this particular background gauge the following diagram commutes,
\begin{displaymath}\label{dia:diagram}
\xymatrix{\textrm{twistor YM} \ar[r]^{\textrm{quant.}}&  \textrm{twistor YM} \ar@{<->}[d]^{Penrose }\\
 \textrm{YM on }R^4 \ar@{<->}[u]^{Penrose } & \textrm{YM on }R^4 \ar@{<-}[l]^{\textrm{quant.}}
}
\end{displaymath}
In a very real sense this result is also expected, since the quantum effective action can always be calculated on space-time for the Chalmers and Siegel action as a functional of the space-time fields $\cA(x)$ and $\cB(x)$. The method outlined in the previous section allows one in principle to lift any functional, so the part which is added in this section is the top arrow. 

Just as in the case described in \cite{us} there is a residual gauge symmetry. When the background field is also in the space-time gauge these are those transformations for which the transformation parameter $\chi$ is a function of $x$ only. Putting the frames back in we arrive at
\begin{equation}\label{eq:resgaugeBG}
\chi(x,\pi) = H \chi(x) H^{-1}
\end{equation}
Of course this can be checked by direct calculation, as these are the transformations for which the gauge-covariant Laplacian vanishes, 
\begin{equation}
(\dbar_0 +A_0)^{\dagger}(\dbar_0 +A_0) \chi = 0.
\end{equation}
What needs checking is whether or not the additional gauge fixing conditions break the carefully preserved background gauge invariance. Below this is verified explicitly by imposing the background version of Lorenz gauge on the background field. 

\subsubsection*{Residual gauge fixing in Lorenz gauge}
The gauge one would like to impose on the quantum field $\ca$ is the usual background field version of the Lorenz gauge,
\begin{equation}
(\partial_\mu + [\cA_\mu,) \ca^{\mu}=0.
\end{equation}
However, it is not immediately obvious this is actually invariant under the background symmetry, the right hand side of (\ref{eq:gaugesym1}). The problem is that the quantum field $\ca_{\alpha \dalpha}(x)$ does not transform nicely under this symmetry, whereas the field $a_\alpha(x,\pi)$ does.  In addition, the weights of the fields are odd: since we want to write down a Lagrange multiplier on space-time, it must be weightless from the point of view of twistor space and this is difficult to achieve with two weight $1$ fields. We must write down a term which is gauge covariant under the background symmetry, a Lorentz scalar, weightless and reduces to the Lorenz gauge if the background field obeys the background gauge. 

From the form of the effective vertices in the Yang-Mills action (\ref{eq:YMbackgroundaction}) it follows that the thing to look for consists of integrations over multiple spheres. This solves the weightless condition. The natural building block is of course the background covariant derivative,
\begin{equation}\label{eq:backgroundgaugederiv}
\dbar_\alpha + [A_{\alpha}(x,\pi_1),
\end{equation}
which transforms in the adjoint at $\pi_1$ if and only if it acts on something which transforms in the adjoint at $\pi_1$. Since this must be true locally, a functional of $a_\alpha(x,\pi_2)$ needs to be constructed which transforms as an adjoint field at $\pi_1$. This is naturally constructed by using link operators in terms of holomorphic frames $(\sim (\dbar + A_0)^{-1})$, see \cite{lionel1})
\begin{equation}
\sim \int_{\CP^1} dk_2 H_1 H_2^{-1} a^\alpha(x,\pi_2)  H_2 H_1^{-1} \frac{\hat{\pi}^{\dalpha}}{\pi \hat{\pi}}. 
\end{equation}
This construction transforms in the adjoint of the background gauge symmetry at $\pi_1$. If we therefore act on it with the background covariant derivative, (\ref{eq:backgroundgaugederiv}) and integrate over $\pi_1$ we obtain the desired background gauge covariant gauge fixing term. Now we can rewrite that term in terms of the field $\ca_{\alpha \dalpha}(x)$
\begin{equation}
\int \frac{\hat{\pi}_1^{\dalpha}}{\pi_1 \hat{\pi_1}} \left(\pi_1^{\dgamma} \frac{\partial}{\partial x^{\alpha \dgamma}} + g [A_{\alpha}(x,\pi_1)\right) \left(H_1 a^{\alpha \dalpha}(x)  H_1^{-1}(x)\right)
\end{equation}
Some further massaging gives
\begin{equation}
\left(\partial_{\mu} + [\cA_{\mu},\right) \ca^{\mu}(x)
\end{equation}
with $\cA$ given by \ref{eq:eqforpentranA}. The point of this exercise is that while the above term does not look background gauge covariant, by its construction it is and in addition it is a Lorentz scalar and weightless as required. Note that this gauge condition involves the metric on $\mathbb{R}^4$. 

\subsection{Renormalization and a conjecture}
Now we have all the ingredients to discuss the renormalization properties of the twistor action in the background space-time gauge. If we also impose the twistor version of the background Lorenz gauge constructed above to fix the residual gauge symmetry, the perturbation series is completely well-defined and by employing dimensional regularization maintains Lorentz invariance and the space-time part of the gauge invariance. A direct consequence of the diagram \ref{dia:diagram} is therefore that the twistor action, in this particular gauge, is therefore as renormalizable as the Chalmers and Siegel action. Since that action reduces to Yang-Mills theory by integrating out the $\cB$ field, it is expected that the Chalmers and Siegel action is renormalizable. From the point of view of the twistor action, the counterterms contain by the lifting formula an infinite sequence of terms. As noted before, this possibility is a consequence of the fact that the mass dimension of the field $A_0$ is zero. However, the calculation in the current section shows that by background gauge invariance, only 3 towers counter-terms are non-trivial, since in the space-time action only 
\begin{equation}
F[\cA]^2 \quad \quad \cB^2 \quad \quad \cB F[\cA]
\end{equation}
counterterms are needed. 

These terms have an intriguing structure from the twistor point of view. Lorentz and space-time gauge invariance only restrict the renormalization $Z$ factors to
\begin{align}
A_0 & \rightarrow Z_{A_0} A^R_0 \nonumber\\
B_0 & \rightarrow  Z_{B_0} B_0 + \left(Z_{BA} F[\cA]_{\dalpha \dbeta}\right) \frac{\hat{\pi}^{\dalpha}\hat{\pi}^{\dbeta}}{(\pi \hat{\pi})^2} \nonumber\\ 
A_\alpha & \rightarrow Z_{A_\alpha} A^R_\alpha \nonumber\\
B_\alpha & \rightarrow Z_{B_\alpha} B^R_\alpha \nonumber\\
g & \rightarrow Z_g g^R
\end{align}
where a possible renormalization of $B_\alpha$ by $\sim \int F_{0\alpha}$ has been discarded since that term vanishes by the constraint. It is easy to see that $Z_{A_0} = Z_{A_\alpha}$, since by gauge invariance, these should appear on an equal footing in $F_{0\alpha}$ and are equal to the usual $Z_A$ renormalization constant. In addition, it is also easy to see that the Feynman rules in the background gauge employed above do not generate contributions to $B^\alpha F_{0\alpha}$, so
\begin{equation}\label{eq:suspi}
Z_{B_\alpha} Z_{A}= 1  
\end{equation}
The usual space-time gauge symmetry argument yields
\begin{equation}\label{eq:renormfacscoupl}
Z_A Z_g =1
\end{equation}
We are therefore left with three independent renormalization constants\footnote{If one wants to fix $\alpha=1$ in a Lorenz-like gauge fixing term $\sim \frac{1}{\alpha} (\partial A)^2$ one needs an extra renormalization constant for this extra coupling, $Z_\alpha$ \cite{abbott}.}
\begin{equation}
Z_A \quad Z_{AB} \quad Z_{B_0}
\end{equation}
Three constants seems superfluous, since Yang-Mills theory itself only needs one. This suggests that there are more relations between the constants, which are not obvious in the chosen set of gauge conditions. Of course, the same question can be asked in the background approach to the Chalmers and Siegel action itself. On a slightly speculative note, we will conjecture one: we suspect that the $BF$ term never diverges. In other words,
\begin{equation}
Z_{B_0} (Z_{A} + Z_{BA})= 1  \quad \quad (\textrm{all loops?})
\end{equation}
This is true at the one-loop level. Furthermore, in the supersymmetric version of the twistor action, the $BF$ term is part of what seems to be an F-term. The underlying observation is that twistor space has a conformal symmetry, so all terms contributing to the local term on twistor space should be 'conformal'. Note that the natural extension of the conjecture is the expectation that \emph{all} local terms on twistor space are in some definite sense protected from quantum corrections. However, at this point this is nothing but a conjecture, which needs further checking. It would of course already be nice to have a definite translation of the usual supersymmetric non-renormalization theorems into twistor Yang-Mills language. 

Using the calculation in the second appendix, it is clear that the twistor Yang-Mills theory has a non-zero $\beta$ function: scale-invariance is broken. Up to a field redefinition (the $Z_{BA}$ term), it can be seen that the $\beta$ function arises by comparing the coefficient in front of the $BF$ and $B^2$ terms. In other words, the beta function is related to the size of the twistor $\CP^1$. This is also expected, as this can be related to the size of the excised twistor line which corresponds to $\infty$. Removing this line breaks the symmetry group of the space from the conformal down to the Poincare group.

\subsubsection*{$\mathcal{N}=4$}
The above analysis does have a nice interpretation in $\mathcal{N}=4$ SYM where a background gauge calculation can be set up just as in this article: if the quantum effects leave supersymmetry unbroken, than the $\beta$ function vanishes. This follows from the observation that in the twistor formulation of $\mathcal{N}=4$ theory \cite{us} $a$ and $b$ are parts of the same super multiplet. They should therefore have the same renormalization constant $Z_A$ if $\mathcal{N}=4$ supersymmetry is unbroken by quantum effects. Therefore by (\ref{eq:suspi})
\begin{equation}
Z_A =1 \quad \quad (\textrm{in } \mathcal{N}=4)
\end{equation}
holds to all orders in perturbation theory. By (\ref{eq:renormfacscoupl})
\begin{equation}
Z_g=1 \quad \quad (\textrm{in } \mathcal{N}=4).
\end{equation}
then follows which in turn implies a perturbatively vanishing $\beta$ function. Of course, the real technical difficulty in this argument lies in proving the assumption that the quantum effects do not break $\mathcal{N}=4$. By the close relation of our techniques to space-time arguments this is fully expected (including the usual caveat about the existence of a supersymmetric regulator), but the background gauge choice employed in this article does break manifest (linear) $\mathcal{N}=4$ supersymmetry. This is probably comparable to the way a Lorenz gauge in real Chern-Simons theory introduces dependency on a metric. 

\section{Towards Yang-Mills S-matrix}
\label{sec:towardssmatrix}
In ordinary Yang-Mills the background field method can be used to calculate S-matrix elements with the background field in a different gauge than the quantum field \cite{abbottgrisaru, abbott}. This observation is based on the fact that the quantum effective action obeys
\begin{equation}\label{eq:abbottargument}
\Gamma[A] = \tilde{\Gamma}[\hat{A},A] \restriction_{\hat{A}=A}
\end{equation}
In ordinary Yang-Mills theory, the left hand side of this equation is the quantum effective action calculated in the background field method, while the right hand side is the effective action of the theory defined by shifting the quantum field, basically undoing (\ref{eq:splitfieldsTWYM}). This equation is derived from the observation that the only difference for the calculation of the effective action between the background field path integral and the usual path integral is the fact that they employ a Legendre transform with different sources: the background field integral has a source $J a$, while the usual path integral has a source $J(\tilde{a}-A)$. The background fields in the right hand side will then appear solely in the gauge fixing part of the action. As the S-matrix is independent of the gauge-fixing functional, it is independent of the background field gauge choice, which can be checked explicitly. This is of course nothing but the statement that physical states correspond to BRST cohomology. We expect that the same proof can in principle be used in the twistor action in a general background gauge. 

Note that it is already clear from the formulae in the previous section that even low-point Green's functions calculated using the background field method contain vertices with a large number of fields in any other gauge than space-time gauge. Two-point functions for instance here calculate already infinite towers of effective vertices! These towers disappear in the case where the background field obeys the space-time gauge condition. It can be taken however as a clear indication that it might be possible to calculate large classes of effective vertices with a few simple diagrams.

\subsection{The background field in CSW gauge}
As indicated before, the background field will be put into CSW gauge, as then at tree level just the MHV formalism is obtained \cite{usII}. Although one could calculate in principle with a Lorenz gauge gauge quantum field, in this case it is more convenient to employ the light-cone formalism for the quantum field. The convenience stems from the fact that the space-time projection of the background field in CSW gauge obeys 
\begin{equation}\label{eq:lightcone}
\eta^{\alpha} \xi^{\dot \alpha} \cA_{\alpha \dot\alpha} =0
\end{equation}
The origin of the arbitrary spinor $\xi$ is elucidated below. This equation follows from equation (\ref{eq:vectcohom}), since evaluating that equation on $\pi=\xi$ gives,
\begin{equation}
\eta^{\alpha} \xi^{\dot \alpha} \cA_{\alpha \dot\alpha} = \eta^{\alpha} \xi^{\dot \alpha} \left(H^{-1}(\xi) \partial_{\alpha \dot{\alpha}}  H(\xi) \right)
\end{equation}
Recall that the holomorphic frames are defined to be the solution to $\dbar_A H|_{p^{-1}(x)}=0$. Solving this equation however requires a boundary condition, in this case the value of the holomorphic frame at a base-point. We pick this point to be $\xi$ and normalise $H(\xi)= 1$. From this short observation equation (\ref{eq:lightcone}) follows. 

The combination $\xi^{\dalpha} \eta^{\alpha}$ forms a null vector in four dimensional space-time, therefore the above result exhibits the close link between the twistor CSW and space-time light-cone gauge. Since the projection of the background field to space-time gauge which couples to the quantum field obeys a light-cone gauge condition, it is natural to impose light-cone gauge on the quantum field as well. In the following for calculational ease the spinor direction indicated by $\eta$ and $\xi$ will be denoted by $1$ and $\dot{1}$ respectively. So for arbitrary spinors $m^{\alpha}$,$n^{\dot\alpha}$,
\begin{equation}
m^{\alpha} = m_1 \eta^{\alpha} +m_2 \hat{\eta}^{\alpha} \quad n^{\dot \alpha} = n_1 \xi^{\dot \alpha} + n_2 \hat{\xi}^{\dot \alpha}
\end{equation}
The light-cone gauge condition on the quantum field therefore becomes
\begin{equation}
\ca_{2 \dot{2}} =0.
\end{equation}
In light-cone coordinates, it is natural to study the physical fields $A_{2 \dot{1}}$ and $A_{1 \dot{2}}$. By equation (\ref{eq:vectcohom}), it follows that the Yang-Mills field on space-time splits into two series of fields on twistor space in the CSW gauge. Roughly we have
\begin{align}
A_{2 \dot{1}} &= a_0 + (a_0 a_0) + \ldots \\
A_{1 \dot{2}} &= b_0 (a_0 + (a_0 a_0) + \ldots),
\end{align}
where the second equality follows by the field equation. 

\subsection{Self-dual sector}
In space-time Yang-Mills it is known that the truncation to just the $BF$ part of the Chalmers and Siegel action generates exactly one series of amplitudes, at one loop: the amplitudes with all helicities equal. This is precisely the series of amplitudes which appear to be projected out in the MHV formalism, and as a first consistency check one would like to know if these are non-zero in this approach. The background coupled action follows by the same method as employed in the previous section, 
\begin{align}\nonumber
S[A,B, a,b] = S[A,B] + &\frac{1}{2} \int d^4x \cB_{\dalpha \dbeta}[\ca] F^{\dalpha \dbeta}[\ca]  + \frac{g}{4} \int d^4\!x  f_{abc} \cB^a_{\dalpha \dbeta} \ca_{\alpha}^{\dalpha,b} \ca^{\alpha \dbeta,c} \nonumber\\
 & + \frac{g}{4} \int d^4\!x  f_{abc} \cb^a_{\dalpha \dbeta} \cA_{\alpha}^{ \{\dalpha,b} \ca^{\alpha \dbeta\},c} .
\end{align}
Here the fields $\cB$ and $\cA$ denote the space-time projection of the twistor fields $A$ and $B$. These fields are put in the CSW gauge. Hence the complete tree-level perturbation theory is automatically trivial, as there are no more vertices whatsoever. As argued above, it is convenient to impose $\ca_{2 \dot{2}} =0$ on the quantum field. Writing out the components of the action yields,
\begin{align}
S[A,B, a,b] =  S[A,B] + & \frac{1}{2} ( \int (- \cb_{\dot1\dot1} \partial_{2 \dot2} \ca_{1\dot2})  + \cb_{\dot2\dot2} (\partial_{1 \dot1} \ca_{2\dot1} - \partial_{2 \dot1} \ca_{1\dot1} + [\ca_{1\dot1},\ca_{2 \dot1}] )  -  \nonumber  \\
& \cb_{\dot1\dot2} \left(\partial_{1 \dot2} \ca_{2\dot1} - \partial_{2 \dot1} \ca_{1\dot2}  - \partial_{2 \dot2} \ca_{1\dot1}  + [\ca_{1\dot2},\ca_{2 \dot1}] \right) + \nonumber \\ 
& \frac{1}{2} (\cB_{\dot2\dot2} [\ca_{1\dot1}, \ca_{2\dot1}] -  \cB_{\dot1\dot2} ([\ca_{1\dot2}, \ca_{2\dot1}]) + \nonumber  \\
& \frac{1}{2} ( \cb_{\dot2\dot2}\left( [\cA_{1\dot1}, \ca_{2\dot1}] - [\cA_{2\dot1}, \ca_{1\dot1}]  \right) -  \cb_{\dot1\dot2} \left([\cA_{1\dot2}, \ca_{2\dot1}] - [\cA_{2\dot1}, \ca_{1\dot2}] \right) ).
\end{align}
Now, following similar steps as in \cite{chalmsieg}, the fields $\cb_{1 \dot1}, \cB_{1 \dot 1}$ can be integrated out. The last field can be integrated out because there are no quantum corrections for this field. Note that this is equivalent to studying the field equation for the twistor field $B_0$ and evaluating the resulting equation on $\pi=\hat{\xi}$. The quantum field will then decouple.  This will set $\hat{\eta}^{\alpha} a_{\alpha}(\hat{\xi}) =0$, which is exactly $\cA_{1 \dot2}$. The obtained solutions are
\begin{equation}
\cA_{1 \dot 2}=0 \quad \ca_{1 \dot 2}=0.
\end{equation}
With these solutions there are no more quantum corrections for $\cB_{\dot1 \dot2}$, and the only place $\cb_{\dot1\dot2}$ features is now in the kinetic term. Therefore these to can be integrated out exactly, and 
\begin{align}
\ca_{1 \dot 1} = p_{1 \dot 2} \bar{\phi} & \ca_{2 \dot 1} = p_{2 \dot 2} \bar{\phi} \nonumber \\
\cA_{1 \dot 1} = p_{1 \dot 2} \bar{\Phi} & \cA_{2 \dot 1} = p_{2 \dot 2} \bar{\Phi}
\end{align}
is obtained. Again, integrating out $B_{\dot1 \dot2}$ can also be performed by studying the field equation for the twistor field $B_0$ and evaluating this on $\pi = a \xi + b \hat{\xi}$. As the left hand side of the field equation is proportional to $\pi^{\dot \alpha} \pi{\dot \beta}$ by the constraint, all components of this equation should vanish separately. 

The obtained solutions can be plugged into the remaining parts of the action and with the definition $\cb_{\dot2 \dot2} = \phi $ the following result
\begin{equation}\label{eq:determinant}
S = S[A,B] + \tr \int \phi \Box \bar{\phi} + \frac{1}{2} ( \phi [\partial_{\alpha \dot 2} \bar{\phi},\partial^{\alpha}_{\dot 2}\bar{\phi}])  + \frac{1}{2} ( \phi [\partial_{\alpha \dot 2} \Phi,\partial^{\alpha}_{\dot 2}\bar{\phi}]) + \frac{1}{2} ( \cB_{\dot2 \dot2} [\partial_{\alpha \dot 2} \bar{\phi},\partial^{\alpha}_{\dot 2}\bar{\phi}])
\end{equation}
is obtained. Here the full background action is retained. In other words the solutions to the field equations are only used for the background fields coupling to the quantum field. This is possible here because the background fields $B$ whose field equations are needed only appear in the classical action. Several things follow from this action. First of all, it can be checked there are no higher than one-loop diagrams. Second, with the background field in CSW gauge the only quantum effects in this theory arise as a field determinant. Thirdly, no loops with external $B's$ will be generated as there are simply no diagrams for pure loops and there are also no tree-level vertices which could give external $B$'s by dressing. If the background field was put in "space-time+light-cone" gauge and treated in the light-cone formalism as well, this is diagrammatically simply what is obtained from using only MHV three (3) vertices as building blocks. In the present setup the calculation is slightly different, as will be illustrated below. 

\subsubsection{The four-point all plus amplitude}
It is instructive to study the four-point all plus amplitude calculated in the above frame-work. Since $\cA_{2 \dot 1}$ can be expanded in terms of $a_0$ twistor fields, there are in principle 3 different contributions. These can be diagrammatically represented by diagrams with the same topology as in the ordinary light-cone case. Within dimensional regularization, the bubble contributions vanish, which leaves the box and the triangles. The box diagram can easily be seen to be equivalent to the lightcone calculation. Therefore we will only need the expansion of $\cA_{2 \dot 1}$ to second order,
\begin{align}
\cA_{2 \dot 1} & \equiv \eta^{\alpha} \hat{\xi}^{\dalpha} \cA_{\alpha \dalpha} = H(\hat{\xi}) (\eta^{\alpha} \hat{\xi}^{\dalpha} \partial_{\alpha\dalpha})H^{-1}(\hat{\xi}) \\
 & = (a_0) + (a_0)^2 + \ldots
\end{align}
One wasy this computation can be done is by an expansion of the frames $H$,
\begin{equation}
H(\hat{\xi}) = 1 + \int_{\CP^1} \frac{a_0(\pi_1)}{\hat{\xi} \pi_1} \frac{\hat{\xi} \xi}{\pi_1 \xi} + \int_{(\CP^1)^2} \frac{a_0(\pi_1)a_0(\pi_2)}{(\hat{\xi} \pi_1)(\pi_1 \pi_2)} \frac{\hat{\xi} \xi}{\pi_2 \xi} + \ldots
\end{equation}
which leads after some algebra to 
\begin{align}\label{eq:ettlemorrisresem}
\cA_{2 \dot 1} & = A_z(p) = \braket{\hat{\xi} \xi} \left(\frac{a_0(q_1)}{\eta \xi q_1} + (\eta \xi p)\frac{a_0(q_1) a_0(q_2)}{(\eta \xi q1)(\eta q1 \eta q2) (\eta \xi q2)}  + \ldots \right).
\end{align}
Here the obvious momentum constraint has been suppressed. Indeed, inserting the external field normalizations and performing all the sphere integrals using the delta functions shows explicitly that the calculation of the triangles are also \emph{exactly} equivalent to the light-cone calculation, diagram by diagram. Hence it follows that the correct scattering amplitudes are reproduced (see e.g. ~\cite{BSTrecent}). Moreover, expressions of the type displayed above (especially \ref{eq:ettlemorrisresem}) are very closely related to the light-cone approach to MHV diagrams advocated by~\cite{mansfield}. Actually the coefficients found by Ettle and Morris~\cite{ettlemorris} can all be reproduced by an extension of the above argument. This will be discussed elsewhere.  

\subsubsection{Towards the off-shell all-plus vertex}
From the setup described above, all the all-plus amplitudes should follow from equation \ref{eq:determinant}. Hence these amplitudes are generated by a determinant. This is in the spirit of \cite{BSTrecent}. One of the aims of the present work was to see if it were possible to calculate complete vertices in the twistor quantum effective action. One of the goals would be to elucidate the twistor structure of the one-loop amplitudes. In particular, as all-plus amplitudes localize on lines in twistor space \cite{wittencachazo}, it is natural to expect in a twistor action formalism that there exists an off-shell \emph{local} vertex in the quantum effective action which reproduces those amplitudes. Remarkably, this is very easy to write down in a CSW gauge as it can be verified that\footnote{This form of the 'all-plus vertex' was first written down by Lionel Mason.}
\begin{equation} 
\Gamma^{(1)}[a_0]   = \int d^4x \int_{(\CP^1)^2}  \partial_{\alpha \dalpha} K_{21} (\partial^{\alpha}_{\dbeta} (a_0)_{(1)} \partial_{\beta}^{\dalpha} K_{12} \partial^{\beta \dbeta} (a_0)_{(2)}	
\end{equation}
reproduces the known answer as a local vertex. Here 
\begin{equation}
K_{12} = \left(\dbar_0 + a_0\right)^{-1}_{12}
\end{equation}
is the full Green's function on the $\CP^1$ sphere. Note this can never be invariant under the full twistor space gauge symmetry as this expression vanishes in space-time gauge. It is therefore perhaps best interpreted as an effective vertex in a gauge-fixed formalism. Unfortunately, apart from the indirect argument that the self-dual Yang-Mills theory generates all all plus amplitudes, we were unable to connect the above vertex to the calculation of the amplitudes directly.

Note however, that the vertex can be promoted to a quantity invariant under space-time gauge transformations. Promoting derivitives to full covariant derivatives using equation \ref{eq:eqforpentranA} would lead to unwanted contributions to the scattering amplitudes, as there would be more $a_0$ fields floating around. In contrast, note that 
\begin{equation}
\tilde{A}_{\alpha \dalpha} = \int \frac{\hat{\pi}_{\dalpha} a_{\alpha}}{ \pi \hat{\pi}}
\end{equation}
transforms like a space-time connection under gauge transformations on twistor space which only depend on spacetime:
\begin{equation}
\delta\tilde{A}_{\alpha \dalpha} = \int \frac{\hat{\pi}_{\dalpha} (\dbar_\alpha f(x) + [a_{\alpha,f(x)])}}{\pi \hat{\pi}} = \partial_{\alpha \dalpha} f + [\tilde{A}_{\alpha \dalpha},f]
\end{equation}
Using this new covariant derivative a vertex invariant under space-time gauge transformations can be devised. We do not know if the extra contributions generated by the 'tilde' covariant derivatives control any scattering amplitude: they most definitely do no generate the four point $+++-$ amplitudes\cite{wittencachazo}. Furthermore, these amplitudes would not be generated within the self-dual theory. It is interesting to note that the structure of the all-plus amplitudes also arises in other contexts~\cite{stieberger}.

\subsection{The full theory}
The same method as in the self-dual sector can in principle be applied to the complete theory. A full treatment will be deferred to future work, but it is possible to predict the result: by this stage it is natural to expect that it amounts to taking light-cone Yang-Mills theory, induce a separation between background and quantum field and apply a twistor lift to the background fields only, keeping these in CSW gauge. The tree level results should be given by just the MHV rules, as we can use the solution to the background field equations only for the background fields coupling to the loops, leaving the tree level action intact. Denoting $+$ and $-$ helicity fields by $\cC$ and $\cbC$ for background and $\cc$ and $\cbc$ for quantum fields we obtain schematically (for the one loop calculation)
\begin{align}\label{eq:edguess}
S_{\textrm{C's}}[\cA,\cB] + S_{\textrm{light-cone}}[\cc, \cbc]  +&  \cc \cbc \cC + \cc \cbc \cbC + \cc \cc \cbC + \cbc \cbc \cC \nonumber \\
+& \cbc \cbc \cC \cC + \cc \cc \cbC \cbC + \cc \cbc \cbC \cC  .
\end{align}
Here $\cC$ and $\cbC$ are given by
\begin{equation}
\cC = \cA_{1 \dot2}(x) \quad \quad \cbC = \cA_{2 \dot1}(x).
\end{equation}
which can be lifted straight to twistor space in the CSW gauge. Interestingly, this decouples the two terms of (\ref{eq:eqforpentranA}). This is also expected, as the linearised $A$ on-shell shows explicitly one term is one and the other the other helicity. Now it can easily be checked that one of these contains at least one $A_{\alpha}$ which gets turned into a $b_0$ by the field equation, while the other only contains $A_0$ fields.  As the action for background fields only generates MHV diagrams at tree level, it is easy to see that the Feynman rules derived for the action above will then generate loop diagrams for both the all $+$ and all $-$ amplitudes at the same time: one follows from combining what are simply the MHV $3$-vertices (the one with the external $A_\alpha$) knotted into a loop, dressed with MHV-trees. The other one then follows as a straight field determinant, without any forestry. Note that this is a neat realisation of both separate scenarios sketched in \cite{BSTrecent} within one framework. Actually it is easy to see that applying Mansfield's canonical field transformation to only the background fields in equation (\ref{eq:edguess}) will realise this scenario. We conjecture this constitutes a full quantum completion of the CSW formalism.

Of course, many things have to be checked far more explicitly than done here. However, as we saw in the previous subsection, at least in principle we can follow the same steps needed to make the formalism run. One large caveat in all of the above is that we have not regulated the action very carefully since the light-cone formalism operates strictly in $4$ dimensions and it would be useful to do this properly. However, these problems are again just space-time ones. One obvious way around them is to write a $4-2\epsilon$ dimensional Yang-Mills action and apply lifting only to the four dimensional degrees of freedom in the spirit of a dimensional reduction. 

\section{Discussion}

In this exploratory article we have shown that there is a class of gauges in which the twistor action formulation of Yang-Mills theory makes sense as a quantum theory in the usual perturbative approach. A partial gauge fixing shows that the quantum effective action of twistor Yang-Mills in this gauge is equivalent to the twistor lift of the quantum effective action of the Chalmers and Siegel action calculated in the background field approach. In particular the divergence structure is the same and in this class of gauges twistor Yang-Mills is as renormalizable as ordinary Yang-Mills in the Chalmers and Siegel formulation. Although it is fully expected that that formulation is equivalent to ordinary Yang-Mills at the quantum level, this is not completely obvious. In particular it would be nice to have a renormalizability proof for the Chalmers and Siegel action, which should be a straightforward extension of results in the literature. In addition a full proof of unitarity would be nice, although again this is expected to hold by the close relationship between the perturbation series of twistor Yang-Mills and the space-time version exposed here. We also have formulated a conjecture on the non-divergence within perturbation theory of local terms in the twistor action based on the twistor structure and a counting argument. This certainly deserves some further study.Unfortunately, apart from the indirect argument that the self-dual Yang-Mills theory generates all all plus amplitudes, we were unable to connect the above vertex to the calculation of the amplitudes directly.

An obvious question remains as to what other gauge choices within the twistor framework are possible and/or interesting. In particular one would like to move away from the space-time oriented background gauge employed here and move toward more twistorial ones. The probably most well-behaved gauge of all for instance, the `generalised Lorenz' gauge ($\dbar^{\dagger} a=0=\dbar^{\dagger} b $), is a natural possibility to consider. Besides choosing a metric on $\CP^3$, this requires however a better understanding of twistor propagators beyond the half-Fourier transform technique employed up to now and in particular their regularization at loop level. It would also still be interesting to find a way to make sense of CSW gauge directly, although there it remains a problem to see how to make systematic sense of the divergence structure. There are indications however that techniques currently being employed in the light-cone approach to MHV diagrams also should be applicable here. 

The background field method as presented in this article can quite readily be employed in any theory for which a twistor action description is available. The general lifting procedure in the form described in this article is actually applicable to large classes of four dimensional gauge theories, amongst which $\mathcal{N}=4$ SYM \cite{us} and the full standard model \cite{usII}. Of course, in the latter case one would also like to have a better understanding of CSW gauge results. This is under study. 

One research direction which might be interesting from this article is the question of twistor geometry within the context of renormalization: is there a natural geometric twistor interpretation of renormalization? In the twistor string context, the Yang-Mills coupling constant is related to the size of the $\CP^1$ instantons in the disconnected prescription. This suggests that the natural direction to look for `renormalization geometry' is actually the non-projective twistor space $\mathbb{C}^4$. However, based on the results in this article it is not yet quite obvious how this might be achieved.

Another interesting avenue to pursue concerns questions of integrability: it is known that the self-dual Yang-Mills equations are in a real sense integrable, see e.g. \cite{lionelwoodhouse}. In fact, the transform to twistor space can in some sense be viewed as an explicit transformation to the free theory (the `action/angle' variables) underlying the integrability. The twistor action approach to full Yang-Mills can then be understood as a perturbation around the self-dual, integrable sector. It is a very interesting question to what extend techniques employed in the study of classical integrable systems may be imported to the full theory. 

However, the most important point to take from this article is that there is a clear indication that at least part of the structure which makes Yang-Mills perturbation theory at tree level so simple extends to loop level in a consistent way. The goal is that exploiting this observation at a much deeper level than here leads to a better understanding of Yang-Mills theory, both perturbatively and non-perturbatively. The study of perturbation theory in this paper is intended to be a stepping stone in that direction, although even in this form it does apparently furnish a completely regularized, well-behaved, off-shell quantum completion of the MHV formalism. This of course needs further work. One of the things to aim for are for instance (analogs of) Witten's twistor space localization arguments as these make precise what kind of `hidden' structure perturbation theory might have. It would be very interesting to see how these arise within the twistor action framework as this seems to be a natural starting point to try to derive them. This carries the great promise of being able to calculate complete generating functionals of loop amplitudes, similar to how MHV amplitudes are used at tree level. We hope to come back to this issue in future work. 

\bigskip
\noindent \textit{Note added in proof}

\noindent After this paper was submitted to the archive, two other preprints appeared which also deal with the problem of quantum completions of the CSW rules for non-supersymmetric Yang-Mills theory~\cite{durhamsouthhampton, queenmaryII}. Both of these propose a more direct solution to the problem and use Mansfield's canonical transformation technique. By the results in this article, especially section \ref{sec:towardssmatrix}, this is \emph{very} closely related to the twistor approach. 
\bigskip

\section*{Acknowledgements}
It is a pleasure to thank Lionel Mason and David Skinner for comments, feedback and collaboration in an early stage of this work. In addition, many thanks to Ruth Britto, Freddy Cachazo, Paul Mansfield, Gabrielle Travaglini and Costas Zoubos for discussions. The work of RB is supported by the European Community through the FP6 Marie Curie RTN {\it ENIGMA} (contract number MRTN-CT-2004-5652).

\appendix
\section{An alternative CSW-like gauge}\label{app:difgaugcond}
In four dimensional quantum field theory one can use 't Hooft's trick to `square' the gauge condition and arrive at a family of gauge fixings. The normal gauge fixing condition and ghost terms follow from the BRST variation of the gauge fixing fermion,
\begin{equation}
S_{\textrm{gaugefix}} = Q_{\textrm{BRST}} \left(\bar{c} G \right)
\end{equation}
where $G$ is the gauge condition to be imposed and $\bar{c}$ is the anti-ghost for which $\delta_{Q} \bar{c} = \lambda$ with $\lambda$ the Lagrange multiplier field. This is replaced by
\begin{equation}
S_{\textrm{gaugefix}} = Q_{\textrm{BRST}} \left (\bar{c} G - \frac{\alpha}{2} \bar{c} \lambda \right).
\end{equation}
Integrating out the Lagrange multiplier from the resulting action gives
\begin{equation}
S_{\textrm{gaugefix}} = \frac{1}{2 \alpha} G^2 + \textrm{ghosts}.
\end{equation}
In the context of the twistor action this is slightly difficult as the gauge conditions have weight: $\eta^{\alpha} a_\alpha$ for instance has weight $1$ if $\eta$ is weightless. Therefore, squaring this condition does not make sense as an integral on the projective space. However, one can easily normalise the gauge fixing vector $\eta$ to have weight $0$:
\begin{equation}
\eta^{\alpha} \rightarrow \frac{\xi^{\dot\alpha} \hat{\pi}_{\dot \alpha}}{\pi \hat{\pi}} \eta^{\alpha}
\end{equation}  
Here $\xi^{\dot\alpha}$ and $\eta^{\alpha}$ are two constant spinors which taken together form a light-like vector on space-time. In the main text of this article $\xi$ arose as the basepoint of the holomorphic frame $H$. With this redefinition of $\eta$ one can now square the gauge fixing condition for $A$. In the limit $\alpha$ is taken to zero this gauge reduces to the CSW gauge employed in \cite{usII}. 

One can now calculate the propagators in this gauge in the standard way. It is expected when also $\eta B =0$ that in the limit $\alpha \rightarrow 0$ the CSW propagators are recovered. This turns out to be untrue, surprisingly. Instead, one needs to impose 
\begin{equation}
\hat{\pi}^{\alpha} \frac{\partial}{\partial x^{\alpha \dot\alpha}} b^{\alpha}=0
\end{equation}
to obtain the same CSW propagators in the limit $\alpha \rightarrow 0$. In addition one obtains
\begin{equation}
:\!B_{\alpha} A_{0}\!: = \frac{\hat{p}_{\alpha}}{(\pi \hat\pi) p^2} 
\end{equation}
At tree level the gauge constructed in this appendix is equivalent to the CSW gauge. At loop level however, there is now a ghost term and a host of non-zero diagrams connected to the vertex in the Chern-Simons part of the action. These seem unattractive however, since their dependence on $\pi$ indicates that propagators connecting \emph{to} a loop can contribute loop momentum factors. In addition, it is hard to see how these contributions could lead to the all plus helicity amplitudes which are missing in the CSW rules.  

%An illustration of this is the diagram indicated below. 
%Working out the integrand we obtain
%\begin{equation}
%2
%\end{equation}
%which does not seem very physical.

\section{Background field calculation for the Chalmers and Siegel action}\label{app:modelcalc}
In this appendix we briefly describe a background field calculation for Yang-Mills theory as formulated as the BF-like Chalmers and Siegel action. This appendix took some inspiration from a calculation in actual BF theory in 4 dimensions from \cite{martellini}, although both our action and the technique applied are different. 

Yang-Mills theory can be formulated as an action with an `auxiliary' self-dual tensor field $cB_{\alpha \beta}$,
\begin{equation}\label{BFaction}
S = \frac{1}{2} \tr \int d^4\!x (\cB^a_{\dalpha \dbeta} F_a^{\dalpha \dbeta} - \frac{1}{2} \cB^a_{\dalpha \dbeta} \cB_a^{\dalpha \dbeta} ),
\end{equation}
where $F$ is the self-dual part of the Yang-Mills curvature:
\begin{equation}
F_{\dalpha \dbeta} = \frac{1}{2} \epsilon^{\beta \alpha} F_{\alpha \dalpha \beta \dbeta}
\end{equation}
In particular, $F_{\dalpha \dbeta}$ is a symmetric tensor. The normalisation of this term is chosen to have
\begin{equation}
F_{\alpha \dalpha \beta \dbeta} = \epsilon_{\alpha \beta} F_{\dalpha \dbeta} + \epsilon_{\dalpha \dbeta} F_{\alpha \beta}
\end{equation}
which is nothing but the usual observation that the curvature splits naturally in self and anti-self-dual parts in spinor coordinates. Note that, in the Abelian case, integrating out $\cA$ yields the electromagnetic dual action. Integrating out $\cB$ from the above action yields the usual Yang-Mills action up to the topological term. More specifically, if just $\cA$ fields are inserted into the path integral, it is clear that the vev calculated in this way will be just the standard Yang-Mills answer (perturbatively). In particular, the $\beta$ function of this theory should be the same. Below we will show this explicitly through the background field method. Note that the path-integral contains an integration over a self-dual \emph{auxiliary} field and a standard gauge field. Hence we do not expect anomalies to arise from the path-integral measure. 

\subsection{Setup}
Split the fields into a background and quantum part
\begin{eqnarray}\label{eq:splitfieldsBF}
\tilde{\cA} = \cA + \ca \nonumber \\
\tilde{\cB} = \cB + \cb
\end{eqnarray}
indicated by capital letters and lower case letters respectively. We are going to calculate the quantum effective action by integrating out $\cb,\ca$ in perturbation theory. As the action is invariant under
\begin{eqnarray}
\tilde{\cA} &\rightarrow& \tilde{\cA} + d_{g \tilde{\cA}} \chi\\
\tilde{\cB} &\rightarrow& \tilde{\cB} +  g [\tilde{\cB}, \chi]
\end{eqnarray}
there are two obvious (disjoint) choices one can make for the symmetry transformations of quantum and background field:
\begin{align}
\cA &\rightarrow \cA + d_{g \cA} \chi & \cA &\rightarrow \cA\nonumber  \\
\cB &\rightarrow \cB + g [\cB, \chi] & \cB &\rightarrow \cB \nonumber \\
\ca &\rightarrow \ca + g [\ca, \chi] & \ca &\rightarrow \ca + d_{g (\cA+ \ca)} \nonumber \chi\\
\cb &\rightarrow \cb + g [\cb, \chi] & \cb &\rightarrow \cb + g [\cB+\cb, \chi]
\end{align}
The objective in a background field calculation is to fix symmetry number two and keep explicit symmetry number one. One might be tempted to impose Lorenz gauge. As the quantum field $\ca$ transforms in the adjoint of the background field transformation, a pure Lorenz gauge would break the background symmetry. This is easily remedied by the background gauge condition:
\begin{equation}\label{eq:backgroundlorenzBF}
\partial_\mu \ca^{\mu} + g [\cA_{\mu}, \ca^{\mu}] =0.
\end{equation}
This gauge condition transforms in the adjoint of the background gauge transformation, so including this condition with a Lagrange multiplier which also transforms in the adjoint, a nice invariant term can be constructed. Following the usual steps we insert the split \ref{eq:splitfieldsBF} and the appropriate gauge fixing and ghost terms into the action, discard the linear terms as only 1 PI diagrams contribute to the quantum effective action and obtain:
\begin{eqnarray}
S[\cA+\ca, \cB+\cb] &=& S[\cA,\cB] + S[\ca,\cb] + \frac{g}{4} \int d^4\!x  f_{abc} \cB^a_{\dalpha \dbeta} \ca_{\alpha}^{\dalpha,b} \ca^{\alpha \dbeta,c}
+ \frac{g}{4} \int d^4\!x  f_{abc} \cb^a_{\dalpha \dbeta} \cA_{\alpha}^{ \{\dalpha,b} \ca^{\alpha \dbeta\},c}\nonumber \\
& &  - \frac{1}{2 \alpha} \left((\partial_\mu \ca^{\mu,a})^2 + 2 g f_{abc} (\partial_\mu \ca^{\mu,a}) \cA_{\nu}^{b} \ca^{\nu,c} + g^2 f_{abc} f^a_{ef} \cA_{\mu}^{b} \ca^{\mu,c} \cA_{\nu}^{e} a^{\nu,f} \right) \nonumber \\
& & + \bar{c} \overleftarrow{D}_{\mu} (\overrightarrow{D}^{\mu}+\ca^{\mu}) c
\end{eqnarray}
Note that the ghosts inherit their symmetry properties from equation \ref{eq:backgroundlorenzBF} by a simple BRST argument: they transform in the adjoint of both symmetry transformations. 

At this point one could either work with the above action directly or one could integrate out the quantum field $\cb_{\dalpha \dbeta}$. The first option, which we also explored, leads generically to more diagrams to be calculated, although results do not change. Also, one would like to stay as close to the original Yang-Mills calculation as possible. Therefore the field $\cb$ will be integrated out using its field equation,
\begin{equation}
\cb_{\dalpha \dbeta}^a = F_{\dalpha \dbeta}^a + \frac{g}{2}  f^{a}_{bc} \cA_{\alpha \{\dalpha}^{b} \ca_{\dbeta\}}^{\alpha,c} 
\end{equation}
Note that $F_{\dalpha \dbeta}$ is a functional of the quantum field $\ca$ only. This equation leads to 
\begin{align}\label{eq:backgroundChalSieg}
S[\cA+\ca, & \cB+\cb] = S[\cA,\cB]+ \int d^4\!x \frac{1}{4} F_{\dalpha \dbeta} F^{\dalpha \dbeta}  - \frac{1}{2 \alpha} (\partial_\mu \ca^\mu)^2 + \bar{c} \overleftarrow{D}_{\mu} \overrightarrow{D+\ca}^{\mu} c  \nonumber \\
&  - \frac{g}{4 \alpha} ( f_{abc} \partial_{\beta \dbeta} \ca^{a \beta \dbeta} \cA_{\alpha \dalpha,b} \ca^{\alpha \dalpha,c}) + \frac{g}{8} \left(f_{abc} \left( \partial_{\beta \{\dalpha} \ca^{\beta,a}_{\dbeta\}} + g f^{a}_{de}\ca^d_{\beta \{\dalpha} \ca^{\beta,e}_{\dbeta\} } \right)  \cA_{\alpha}^{\{\dalpha,b} \ca^{\dbeta\} \alpha,c} \right) \nonumber \\
&   - \frac{g^2}{8 \alpha}   f_{abc} f^a_{ef} \cA_{\alpha \dalpha}^{b} \ca^{\alpha \dalpha,c} \cA_{\beta \dbeta}^{e} \ca^{\beta \dbeta,f}  + \frac{g^2}{16} f_{abc} f^{a}_{ef}  \cA_{\alpha}^{\{\dalpha,b} \ca^{\dbeta\} \alpha,c} \cA_{\beta \{\dalpha}^{e} \ca_{\dbeta\}}^{\beta,f} \\
&  + \frac{g}{4} \int d^4\!x  f_{abc} \cB^a_{\dalpha \dbeta} \ca_{\alpha}^{\dalpha,b} \ca^{\alpha \dbeta,c}
 \nonumber 
\end{align}
The above action admits an intriguing simplification beyond the simple form of the propagator if $\alpha=1$. In that case the terms quadratic in quantum fields in the second and third lines combine to form
\begin{equation}
\frac{g}{2} f_{abc} \partial_{\beta \dalpha} \ca^{\beta}_{\dbeta,a} \cA_{\alpha}^{\dalpha,b} \ca^{\dbeta \alpha,c} + \frac{g^2}{4} f_{abc} f^{a}_{ef}  \cA_{\alpha}^{\dalpha,b} \ca^{\alpha \dbeta,c} \cA_{\beta \dalpha}^{e} \ca_{\dbeta}^{\beta,f}
\end{equation} 
\noindent which can be proven by decomposing the primed tensor structure in this term into symmetric and antisymmetric parts. Hence we will fix $\alpha=1$ in the following. Note that decomposing the above expression in symmetric and antisymmetric parts in the \emph{unprimed} tensor structure leads to the type of terms which might be derived from the anti-Chalmers and Siegel action (the parity conjugate action),
\begin{equation}
S_{\textrm{anti-Chalmers and Siegel}} = \frac{1}{2} \tr \int d^4\!x (\cB^a_{\alpha \beta} F_a^{\alpha \beta} - \frac{1}{2} \cB^a_{\alpha \beta} \cB_a^{\alpha \beta} ),
\end{equation}
treated in the background field method. It would be interesting to understand this connection further as parity invariance is obscured in the Chalmers and Siegel action. The kinetic term $F_{\dalpha \dbeta} F^{\dalpha \dbeta}$ term can be written as a sum of (minus) the usual Yang-Mills action and a topological term. It therefore follows that perturbatively the quantum field $\ca$ can be treated as a standard $d$-dimensional vector field in dimensional regularization with the 't Hooft-Veltman prescription which only continues fields inside the loop. 

\subsection{Self-energies}
Using \ref{eq:backgroundChalSieg} we can calculate the self-energies of the fields in the theory, 
\begin{equation}
<\cA \cA> \quad <\cA \cB> \quad <\cB \cB> \quad <\bar{c} c>
\end{equation}
As this does not affect the $\beta$ function we will ignore the ghost self-energy here. There is a slight irritation with the normalisation $g_{\mu\nu} = \frac{1}{2} \epsilon_{\alpha \beta} \epsilon_{\dalpha \dbeta}$ in the calculations below: when the quantum fields are rewritten as actual Lorentz-vector fields contracted into a space-time tensor, one picks up a factor of $2$. Note that this rewriting has to be performed in order to employ dimensional regularization. 

\subsubsection*{$\cB \cB$}
By straightforward calculation
\begin{equation}
<\cB \cB> = -\frac{1}{2}\frac{g^2 C_A}{(4 \pi)^{d/2}} \frac{\Gamma[1-\epsilon]^2}{\Gamma[2-2\epsilon]} \Gamma[\epsilon] \int \frac{d^4\!q}{(2 \pi)^{d}} \cB_{\dalpha \dbeta}^a(q) \cB^{\dalpha \dbeta a}(-q) \left(\frac{1}{q^2} \right)^{\epsilon}
\end{equation}
is obtained. This can be expanded as
\begin{equation}
<\cB \cB> =  -\frac{1}{2} \frac{g^2 C_A}{(4 \pi)^{2}} (\frac{1}{\epsilon})\int \frac{d^4\!q}{(2 \pi)^{d}} \cB_{\dalpha \dbeta}^a(q) \cB^{\dalpha \dbeta a}(-q) + \mathcal{O}(\epsilon^0)
\end{equation}

\subsubsection*{$\cB \cA$}
This self-energy vanishes. This is a consequence of the fact that $\cB$ is a symmetric tensor, as the only contribution to this self-energy which is not a tadpole is
\begin{equation}
\sim  f_{abc}  f_{def}  B^a_{\dalpha \dbeta} A_{\alpha}^{\dgamma,e} \int d^4p < :a_{\alpha}^{\dalpha,b} a^{\alpha \dbeta,c}: :\partial_{\beta \dgamma} a^{\beta}_{\dot{\delta},d}  a^{\dot{\delta} \alpha,f}: > .
\end{equation}
Note that one type of vertex has primed indices contracted, while the other has unprimed indices contracted. Working out the contractions with the usual $\alpha=1$ propagator, it quickly emerges that a consequence of this is that the (four dimensional) tensor structure of the primed indices is $\sim \epsilon^{\dalpha \dbeta}$. Contracted into $\cB_{\dalpha \dbeta}$ this gives zero. It is obvious a similar argument applies to many terms in other one-(and higher-)loop diagrams, and at one loop there seem to be no diagrams which contain one $B$ and the rest $A's$ on the external lines. However, there might be sub-leading contributions in $\epsilon$ as for this reasoning to work we must be able to treat the quantum field $\ca$ as a pure $4$ dimensional object which clashes with dimensional regularization: one should do all index contractions on the quantum fields in $d$ dimensions and then couple to the four dimensional background fields. 

The above observation does show that in some sense $\cB$ and $\cA$ seem to couple to disjoint gluons in the loop and it would be very interesting to make this sense precise. An argument in favour of a `decoupling' scenario is the fact that by writing the kinetic term as 
\begin{equation}
F^2 = - \frac{1}{2} \left(F_+^2 + F_{-}^2\right)
\end{equation}
one can argue (at one loop) that the term with the background field $B$ might be part of a determinant which only features self-dual connection, while the terms with the gauge field $A$ might be part of a determinant with the anti-self-dual connection. Of course, this does not touch upon ghost terms and higher loop effects, but it does seem suggestive. 

\subsubsection*{$\cA \cA$}
There are two contributions, one with a gluon in the loop and one with a ghost loop. Note that in the usual Yang-Mills theory these are actually the only two diagram topologies contributing to the $\beta$ function calculation at one loop. The ghost loop is the same as in Yang-Mills and gives
\begin{equation}
<\cA \cA>_{\textrm{ghost}} =   -\frac{g^2 C_A}{(4 \pi)^{d/2}} \frac{1}{\epsilon-1}\Gamma[\epsilon] \frac{\Gamma[2-\epsilon]^2}{\Gamma[4-2 \epsilon]} \int \frac{d^4\!q}{(2\pi)^4} \cA_{\mu}^a(q) (q^2 g^{\mu \nu} - q^{\mu} q^{\nu}) \cA_{\nu}^a(-q)  \left(\frac{1}{q^2} \right)^{\epsilon}
\end{equation}
which yields
\begin{equation}
<\cA \cA>_{\textrm{ghost}} = \frac{1}{6} \frac{g^2 C_A}{(4 \pi)^{2}} \frac{1}{\epsilon} \int \frac{d^4\!q}{(2\pi)^4} \cA_{\mu}^a(q) (q^2 g^{\mu \nu} - q^{\mu} q^{\nu}) \cA_{\nu}^a(-q) + \ldots (\epsilon^0)
\end{equation}

\noindent The gluon loop yields
\begin{equation}
<\cA \cA>_{\textrm{gluon}} =  4 \frac{g^2 C_A}{(4 \pi)^{d/2}} \frac{\Gamma[2-\epsilon]^2}{\Gamma[4-2\epsilon]} \Gamma[\epsilon] \int \frac{d^4\!q}{(2\pi)^4}  \cA_{\mu}^a(q) \left( q^2 g^{\mu \nu} - q^{\mu} q^{\nu}\right) \cA_{\nu}^a(-q) \left(\frac{1}{Q^2} \right)^{\epsilon}
\end{equation}
\noindent which can be expanded as
\begin{equation}
<\cA \cA>_{\textrm{gluon}} = \frac{2}{3} \frac{g^2 C_A}{(4 \pi)^{2}} \int \frac{d^4\!q}{(2\pi)^4} \cA_{\mu}^a(q) (q^2 g^{\mu \nu} - q^{\mu} q^{\nu}) \cA_{\nu}^a(-q) + \ldots (\epsilon^0)
\end{equation}

\subsection{$\beta$ function}
In order to renormalize the theory, renormalization $Z$ factors for the different background fields in the problem are introduced which preserve the Lorentz and gauge symmetries. The latter are preserved as it is known that the regularization procedure employed here does not break gauge invariance and we have set up our calculation explicitly to preserve it. In principle one could introduce renormalization factors for the quantum fields as well, but this never matters as those $Z$ factors cancel between propagators and vertices. As explained in \cite{abbott} the only exception to this is a possible renormalization of the gauge fixing parameter $\alpha$, but this will only contribute at higher loop orders. Since this appendix is only concerned with a one loop calculation, this will be ignored. We get
\begin{align}\label{eq:renormfacs}
\cA^0 &= Z_\cA \cA^R \nonumber \\
\cB^0 &= Z_\cB \cB^R + Z_{\cB\cA} F^R \nonumber \\
g^0 &= Z_g g^R
\end{align}
The symmetries permit an extra field mixing renormalization term for $\cB$ since $F$, like $\cB$, transforms in the adjoint of the gauge group and is a (self-dual) 2-form. In the following the extra superscript $R$ will be suppressed in order to streamline the presentation. Plugging \ref{eq:renormfacs} into the classical action we obtain
\begin{equation}
S_{\textrm{ren}} =  \frac{1}{2} \int d^4\!x (Z_\cA Z_\cB - Z_\cB Z_{\cB\cA}) \cB_{\dalpha \dbeta} F^{\dalpha \dbeta} +  \frac{1}{2} (Z_\cA Z_{\cB\cA} - \frac{1}{2} Z^2_{\cB\cA}) F_{\dalpha \dbeta} F^{\dalpha \dbeta}  - \frac{Z_\cB^2}{4} \cB_{\dalpha \dbeta} \cB^{\dalpha \dbeta}
\end{equation}
Here the field $F$ is the self-dual part of the usual curvature tensor, which is renormalised to
\begin{equation}
\sim d_{[\mu} \cA_{\nu]} + Z_g Z_\cA [\cA_{\mu},\cA_{\nu}]
\end{equation}
In order for the renormalized action to be background gauge invariant,
\begin{equation}
Z_g Z_\cA =1
\end{equation}
must hold. Calculating $Z_\cA$ therefore determines $Z_g$, which can be used to determine the $\beta$ function through \cite{abbott}:
\begin{equation}\label{eq:calcbetafromZ}
\beta(g) =  - g^2 \frac{\partial}{\partial g} Z^1_{A}
\end{equation} 
\noindent Here $Z^1_{\cA}$ is the residue at the pole $\frac{1}{\epsilon}$ in the Laurent expansion of $Z_\cA$. In the background field formalism one can therefore calculate the beta function from a self-energy calculation which is usually much simpler than the 3 point function one needs to calculate otherwise. 

The $Z$-factors can be used to cancel the divergences calculated above. Note that from the divergent parts of the self-energy calculations we get equations which yield (temporarily restoring the loop counting parameter)
\begin{eqnarray}
Z_\cA = 1 + \frac{11}{6} \frac{g^2 C_A}{(4 \pi^2) \epsilon} (\hbar) + \textrm{h.o.}\\
Z_{\cB\cA} =  \frac{5}{6} \frac{g^2 C_A}{(4 \pi^2) \epsilon}(\hbar) + \textrm{h.o.}\\
Z_\cB = 1 - \frac{g^2 C_A}{(4 \pi^2) \epsilon} (\hbar) +\textrm{h.o.}
\end{eqnarray}
which in turn yields the well-known one-loop Yang-Mills $\beta$ function through \ref{eq:calcbetafromZ}
\begin{equation}
\beta(g) = - \frac{11}{3} \frac{g^3 C_A}{(4 \pi^2)}
\end{equation}

\bibliographystyle{JHEP}
\bibliography{lib}

\end{document}